\documentclass[conference]{IEEEtran}
\IEEEoverridecommandlockouts

\usepackage{svg}
\usepackage{booktabs}
\usepackage{multirow}
\usepackage{makecell}
\usepackage[caption=false,font=normalsize,labelfont=sf,textfont=sf]{subfig}

\usepackage{cite}
\usepackage{amsmath,amssymb,amsfonts}
\usepackage{algorithmic}
\usepackage{graphicx}
\usepackage{textcomp}
\usepackage{xcolor}
\def\BibTeX{{\rm B\kern-.05em{\sc i\kern-.025em b}\kern-.08em
    T\kern-.1667em\lower.7ex\hbox{E}\kern-.125emX}}
\begin{document}

\title{A Measurement-Calibrated AI-Assisted Digital Twin for Terahertz Wireless Data Centers
}

\author{\IEEEauthorblockN{
        Mingjie Zhu, Yejian Lyu, Ziming Yu, and Chong Han
    }
}

\maketitle

\begin{abstract}
Terahertz (THz) wireless communication has emerged as a promising solution for future data center interconnects; however, accurate channel characterization and system-level performance evaluation in complex indoor environments remain challenging. In this work, a measurement-calibrated AI-assisted digital twin (DT) framework is developed for THz wireless data centers by tightly integrating channel measurements, ray-tracing (RT), and implicit neural field (INF) modeling. Specifically, channel measurements are first conducted using a vector network analyzer at 300 GHz under both line-of-sight (LoS) and non-line-of-sight (NLoS) scenarios.
RT simulations performed on the Sionna platform capture the dominant multipath structures and show good consistency with measured results. Building upon measurement and RT data, an RT-conditioned INF is developed to construct a continuous radio-frequency (RF) field representation, enabling accurate prediction in RT-missing NLoS regions. Relied on the developed DT, coverage and interference analyses are carried out for access point and rack-level deployments, demonstrating the effectiveness of the proposed DT for network planning and architectural optimization in THz wireless data centers.
\end{abstract}

\section{Introduction}
Modern data centers play a critical role in supporting cloud computing, artificial intelligence (AI), and large-scale distributed storage systems~\cite{7393451}.
With the rapid growth of data-intensive applications, the internal traffic of data centers has increased dramatically, calling for a more flexible and low-latency architecture~\cite{8367741}.
Terahertz (THz) wireless data center (WDC) has emerged as a promising connection architecture for the next-generation AI clusters~\cite{han2026wirescantupreconfigurable}, due to its ultra-wide bandwidth, highly directional transmission, and short-range point-to-point connectivity~\cite{9766110}.
These characteristics make THz wireless links capable of supporting extremely high data rates with low latency and high energy efficiency, while also enabling dynamic and reconfigurable network topologies.
However, the practical deployment and operation of THz wireless data centers face substantial challenges. THz propagation is highly sensitive to environmental dynamics, which is prevalent in realistic data center scenarios. As a result, reliable network planning, interference management, and real-time resource allocation cannot rely solely on static models or offline simulations. This motivates the need for an accurate and adaptive digital twin (DT)~\cite{9120192, 11261676} that can faithfully represent the wireless environment and support system-level decision-making.

A DT for wireless data centers serves as a virtual replica of the physical environment, enabling continuous monitoring, performance prediction, and network optimization~\cite{2025arXiv251107789Z, 10622627}. In~\cite{2025arXiv251107789Z}, a DT of an in-vehicle scenario is established via ray-tracing (RT). In~\cite{10622627}, a multi-layer DT is proposed, but limited to the network layer, not mentioning the physical channel.
However, existing studies on wireless DTs primarily rely on RT or empirical propagation models.
Beyond DT-oriented studies, extensive efforts have been devoted to channel and link-quality prediction using environment-aware or data-driven approaches, e.g., supervised learning methods~\cite{10682525, 9354041, 10.1145/3551660.3560912}.
To obtain trustworthy data, to find out the propagation characteristics in the data center scenario, a growing body of literature has investigated channel measurements in data center environments~\cite{9039668, 10001052, 10609428}, providing valuable empirical insights into path loss, delay spread, and angular characteristics. Nevertheless, these measurement campaigns are typically conducted for channel characterization purposes only and are rarely integrated into a DT framework or exploited for system-level analysis, such as coverage evaluation and interference assessment. Consequently, there remains a clear gap between channel measurements and operational digital twins for wireless data centers, especially in the THz band.

In this work, we aim to bridge this gap by developing a measurement-calibrated DT for THz wireless data centers. We first conduct a vector-network-analyzer (VNA)-based channel measurement campaign in a data center environment to obtain propagation characteristics, which serve as the physical foundation. Building upon this measurement data, we propose an AI-enhanced DT based on implicit neural fields (INF). By modeling the wireless channel as a continuous function of spatial coordinates and RT-derived features, the proposed INF-based twin complements the RT modeling and effectively captures the propagation behaviors that are difficult to model analytically. 
To demonstrate the effectiveness of the proposed digital twin, we leverage it to perform system-level analyses, including coverage mapping and interference evaluation in a THz wireless data center scenario. These results illustrate the potential of the proposed framework to support network planning and performance assessment in future THz-enabled data centers.

The remainder of the paper is organized as follows. The architecture of our proposed DT is introduced in section~\ref{section:dt}. The data processing methods are explained in section~\ref{section: dp}. The case study results of the data center scenario are illustrated in section~\ref{section: cs}. Finally, Section~\ref{section: con} draws the conclusion.

\section{Digital Twin Architecture}
\label{section:dt}
In this section, a DT framework is introduced to establish a connection between the physical measurement environment and its digital counterpart. The physical twin includes the measurement setups, while the digital twin includes two phases, measurement-calibrated RT twin and INF-based AI twin, respectively.
\subsection{Measurement-based Physical Twin}
The measurement-based physical twin provides fundamental physical information for DT construction. Specifically, VNA-based channel measurements are performed to capture channel characteristics in the data center environment, including MPCs, path loss, and delay spread, which serve as the cornerstone for constructing the DT. 

\begin{figure}
    \centering
    \subfloat[]{
        \includegraphics[width=0.79\linewidth]{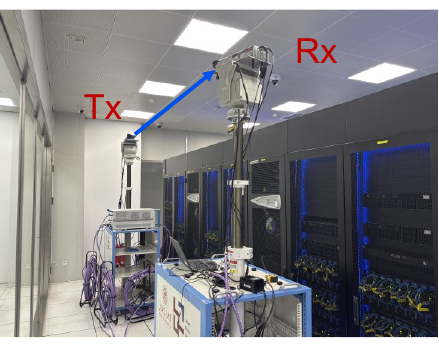}
    }
    \hfill
    \subfloat[]{
        \includegraphics[width=0.78\linewidth]{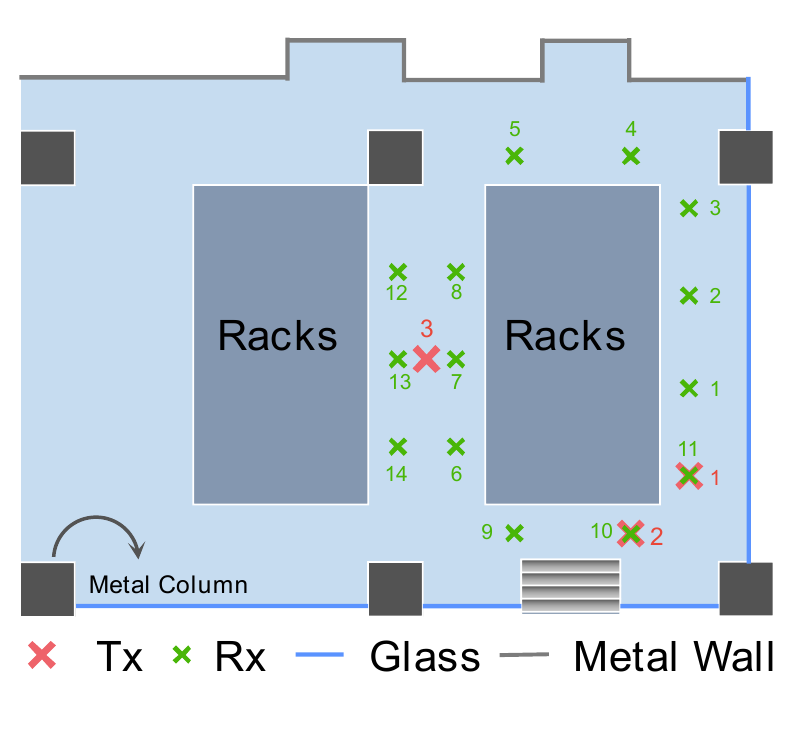}
    }
    \caption{Data center scenario scheme.}
    \label{fig:scheme}
\end{figure}

Fig.~\ref{fig:scheme}(a) shows a photograph of the measurement setup, where a VNA-based channel sounder is used to capture the channel frequency responses (CFRs) inside the data center scenario. The measurement is conducted in the $290$-$310$~GHz range. The broad bandwidth of $20$~GHz brings a delay resolution of $0.05$~ns, ensuring the distance accuracy of $1.5$~cm. The transmitter (Tx) is fixed, while a directional scanning scheme is employed at the receiver (Rx), covering an azimuthal range of $0^{\circ}$ to $355^{\circ}$ and a zenith range of $-20^{\circ}$ to $20^{\circ}$, both with a rotation step of $5^{\circ}$. 
Before the measurements, back-to-back calibration is performed to eliminate system responses. 
The measurement configurations are summarized in Table~\ref{tab:cfg}.

\begin{table}[t]
    \caption{Measurement Configuration.}
    \label{tab:cfg}
    \centering
    \begin{tabular}{cc}
    \toprule
        Parameter & Value \\
    \midrule
        Frequancy Range & $290$-$310$~GHz \\
        Bandwidth & $20$~GHz \\
        Frequqncy Point & $2001$ \\
        IF Bandwidth & $1$~kHz \\
        Transmitted Power & $10$~dBm  \\
        Tx \& Rx Attenna Type & Horn \\
        Tx Height (Rack-to-Rack Case) & $2.4$m \\
        Tx Height (AP Case) & $2.7$m \\
        Rx Height (Rack-to-Rack Case) & $2.4$m \\
        Rx Height (NLoS Case) & $1.7$m \\
        Azimuth Rotation Step & $5^{\circ}$ \\
        Azimuth Rotation  Range & $[0^{\circ}, 355^{\circ}]$ \\
        Zenith Rotation Step & $10^{\circ}$ \\
        Zenith Rotation Range & $[-20^{\circ}, 20^{\circ}]$ \\
        Attenna Gain & 26 dBi \\
        Half Power Beamwidth & $8^\circ$ in horizontal / $6^\circ$ in vertical \\
        
    \bottomrule
    \end{tabular}
\end{table}

As illustrated in Fig.~\ref{fig:scheme} (b), the measurements covered a total of four cases in the data center, corresponding to line-of-sight (LoS) and non-line-of-sight (NLoS) conditions for both Rack-to-Rack and AP-to-User links. A total of 29 Tx-Rx locations are collected. 
Tx~$1$ is positioned alongside the rack, with 9 LoS and 3 NLoS Rxs alongside the rack, simulating the rack-to-rack communication. Tx~$2$ is positioned in front of the rack, with 5 Rxs (Rx~$4-8$) alongside the rack, representing the rack-to-rack communication in the other direction. Tx~$3$ simulates an access point (AP), located at the center of the ceiling, with 9 LoS and 3 NLoS Rxs alongside the rack, simulating AP-to-rack communication.

\subsection{AI-assisted Digital Twin}
\begin{figure*}
    \centering
    \includegraphics[width=0.8\linewidth]{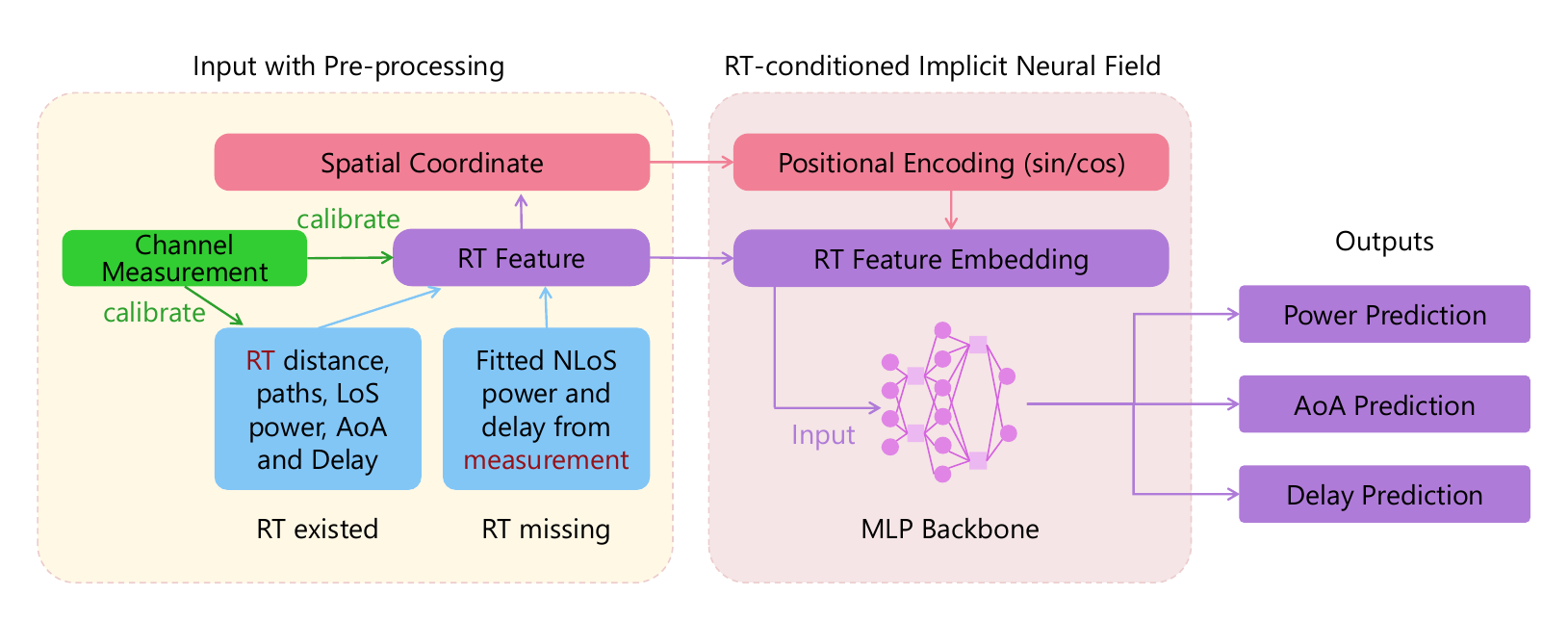}
    \caption{RT-conditioned implicit neural field.}
    \label{fig:inf}
\end{figure*}

An AI-assisted DT is developed to construct a spatially continuous and physics-consistent representation of THz wireless propagation in the data center. 
Specifically, conventional RT is first performed using the open-source simulator Sionna, which provides deterministic propagation paths and coarse multipath characteristics derived from the reconstructed scene geometry and material properties.
It is noteworthy that THz material database is constructed in~\cite{11153056}, which is used in this paper for digital map construction while geometry is obtained by empirical measurement.
Measurement data from the physical twin is employed to calibrate the RT results, constructing the RT twin.  These measurement-calibrated RT results serve as a structural propagation prior, capturing the dominant LoS and specular reflection components. To overcome the limitations of RT in complex non-line-of-sight (NLoS) regions and sparsely measured areas, an INF model~\cite{10.1145/3503250} is employed.
An INF represents a continuous physical field by parameterizing it as a coordinate-based neural network, which directly maps low-dimensional inputs to target physical quantities, especially for modeling spatially smooth yet highly non-linear radio propagation characteristics.
In this work, the INF learns a continuous mapping from spatial coordinates and RT features to key channel attributes, including received power, angles of arrival, and propagation delays, as illustrated in Fig.~\ref{fig:inf}.
By conditioning the INF on the RT twin, the proposed AI twin can refine deterministic predictions and infer propagation characteristics at arbitrary spatial locations, effectively bridging sparse measurement anchors and continuous spatial representations. This RT-based IFN paradigm enables the generation of high-resolution radio maps and multipath profiles across the entire data center, even in regions where direct measurements are unavailable. The accuracy and robustness of the AI-assisted DT are systematically validated through comparisons with real measurement data.

Based on the constructed DT, system-level performance analysis is further enabled. Coverage and interference characteristics between rack-to-rack links as well as between APs and racks are evaluated using the spatially continuous channel representation. These metrics provide real-time and fine-grained insights into network performance, forming a foundation for higher-level optimization and control tasks in data center wireless systems.

\section{Data Processing for DT}
\label{section: dp}
\subsection{Signal Model}

The received THz channel at the Rx can be represented in the delay–angle domain as a superposition of multiple multipath components (MPCs), expressed as
\begin{equation}
\begin{split}
     h(\tau, \theta, \varphi) & = \sum_{\ell=1}^{L} \alpha_{\ell} \delta(\tau - \tau_{\ell}) \cdot \delta(\theta - \theta_{\ell}) \cdot\delta(\varphi - \varphi_{\ell}),
\end{split}
\end{equation}
where $\alpha_{\ell}$, $\theta_{\ell}$, $\varphi_{\ell}$, and $\tau_{\ell}$ stands for the complex amplitude, azimuth and zenith angle of arrival, and delay of arrival of the $\ell^{\mathrm{th}}$ path among all the MPCs with a total number of $L$, respectively.

Based on the measured channel frequency responses and the directional scanning procedure, the parameter set of each MPC is estimated by a local maximum method~\cite{2025arXiv251107789Z} as
\begin{equation}
    \mathbf{\Theta}_{\ell}=\{\tau_{\ell},~\theta_{\ell},~\varphi_{\ell},~\alpha_{\ell}\}.
\end{equation}

The total received power, or equivalently the path loss, is obtained by coherently aggregating the power contributions of all resolvable MPCs, i.e.,
\begin{equation}
    \mathrm{PL}=\sum_{\ell=1}^{L} |\alpha_{\ell}|^2.
\end{equation}

To model large-scale attenuation trends and provide a compact representation for NLoS extrapolation, the measured path loss is further approximated using the ABG model
\begin{equation}
    \mathrm{PL}_{\mathrm{ABG}}=\alpha \log_{10}d+\beta,
    \label{equation: pl}
\end{equation}
where $d$ denotes the Tx–Rx distance, and $\alpha$ and $\beta$ are model parameters obtained via regression over the measurement data. This model is later used to regularize propagation characteristics in regions with sparse or missing RT results.

\subsection{RT-conditioned Implicit Neural Field}
To compensate for the sensitivity limitations of RT, an INF is introduced to learn a continuous representation of the radio channel conditioned on both spatial coordinates and RT-derived features.
The learning objective of the INF can be formulated as
\begin{equation}
    \min \limits_{f_{\theta}}|f_{\theta}-h(\theta)|^2,
\end{equation}
where $f_{\theta}$ denotes a neural network parameterized by $\theta$, and $h(\theta)$ represents the ground-truth channel characteristics obtained from measurements. In practice, the INF learns a mapping from spatial locations and RT features to key channel parameters, expressed as
\begin{equation}
    f_{\phi}(\mathbf{x},\mathrm{RT}) \rightarrow (P,~\tau,~\theta,~\varphi),
\end{equation}
where $\mathbf{x}=(x,y,z)$ is the Rx coordinate and $P$, $\tau$, $\theta$, $\varphi$ denote the received power, delay, azimuth angle, and zenith angle, respectively. And the RT feature vector used to condition the INF is defined as
\begin{equation}
    \mathrm{RT}=[d,~P_{\mathrm{LoS}},~\theta_{\mathrm{LoS}},~\tau_{\mathrm{LoS}},~\varphi_{\mathrm{LoS}},~N_{\mathrm{paths}}].
\end{equation}
where $d$ is the Tx–Rx distance, and $N_{\mathrm{paths}}$ is the number of traced paths. It is worth noting that only a compact set of dominant RT features is used as input, rather than all MPCs, in order to avoid excessive network complexity and ensure scalability of the INF. When RT is missing in some area, the fitted NLoS path loss in ~(\ref{equation: pl}) will replace the missing RT input.

To stabilize training and balance the contributions of different physical quantities, normalization is applied as
\begin{align}
    P^{'} & =\frac{P-\mathbb{E}(P)}{\mathbb{D}(P)}, ~
    {\tau}^{'}  =\frac{{\tau}-\mathbb{E}({\tau})}{\mathbb{D}({\tau})}, ~
    \Omega^{'}  = \frac{\Omega}{360^{\circ}},
\end{align}
where $\Omega=(\theta,~\varphi)$ stands for the three-dimensional angle of arrival. The overall training loss is defined as

\begin{equation}
    \mathcal{L}=||P-\hat{P}||^2+||{\tau}-\hat{{\tau}}||^2+||{\Omega}-\hat{\Omega}||^2,
\end{equation}
which jointly enforces accurate reconstruction of power, delay, and angular characteristics. Through this learning process, the INF refines and complements RT predictions, enabling spatially continuous RF field reconstruction across the data center.

To improve fidelity, the RT outputs are calibrated using channel measurements from the physical twin by aligning dominant MPCs in terms of delay-angle consistency. Specifically, for each measured path $\ell$, the closest RT path is identified via
\begin{equation}
k^{\star}=\arg\min_k \left(
w_{\tau}|\tau_{\ell}-\hat{\tau}k|^2
+w_{\theta}|\theta_{\ell}-\hat{\theta}k|^2
+w_{\varphi}|\varphi_{\ell}-\hat{\varphi}k|^2
\right),
\end{equation}
where $w{\tau}$, $w_{\theta}$, and $w_{\varphi}$ are weighting factors that balance delay and angular domains.
Based on the matched paths, a calibration factor is applied to compensate for modeling errors in propagation loss,
\begin{equation}
\Delta P_k = P_{\ell}-\hat{P}_{k^{\star}},
\end{equation}
which accounts for unmodeled effects such as material uncertainties, diffuse scattering, and partial blockage. After calibration, the RT results provide a measurement-consistent multipath skeleton, preserving the physical structure of propagation while reducing systematic bias.

\subsection{Interference and Coverage}
Based on the constructed digital twin, system-level performance metrics such as interference and coverage~\cite{9247469} can be efficiently evaluated. The signal-to-interference-plus-noise ratio (SINR) at a Rx is given by
\begin{equation}
\begin{split}
    \mathrm{SINR} = \frac{S}{I + P_N} = \frac{P_tG_0K_ug(u)}{\sum_{i\in{\Phi_0/AP_0}}P_tG_iK_{x_i}g(x_{i}) + N_0B},
\end{split}
\label{equation: SINR}
\end{equation}
where $P_t$ is the transmit power, $G_0$ and $G_i$ denote the antenna gains of the desired and interfering links, $K_u$ and $K_{x_i}$ capture path loss and shadowing effects obtained from the digital twin, $g(\cdot)$ represents small-scale fading, $N_0$ is the noise spectral density, and $B$ is the system bandwidth.
Coverage probability is defined as the probability that the received SINR exceeds a predefined threshold $T$, i.e.,
\begin{equation} \begin{split} P_c(T)=\iiint_{V}(p_L(u)P_{c,L}(T)+p_N(u)P_{c,N}(T))f(v)\mathrm{d}V, \end{split} \label{equation: CP} \end{equation}
where $f(v)$ is the probability distribution of the Rxs, and $P_{c,L}$ is the conditional coverage probability of LoS case where the 3D distance is given, and can be calculated as \begin{equation} \begin{split} P_{c,L}(T) = \mathbb{P}(\mathrm{SINR}>T|u)=\mathbb{P}(\frac{P_tG_0K_ug(u)}{N_0B}>T), \end{split} \end{equation}
and correspondingly,$ P_{c,N}$ is the conditional coverage probability of NLoS case.
In the single-transmitter deterministic case, $K_u$ is given by the digital twin, and coverage probability can be obtained subsequently as
\begin{equation} \begin{split} P_{c,L}(T) = (K_u>\frac{N_0BT}{G_0P_t}). \end{split} \end{equation}

\section{Performance Evaluation and Analysis}
\label{section: cs}
\subsection{Measurement Results of THz Data Center}

\begin{figure}
    \centering
    \subfloat[]{
        \includegraphics[width=0.87\linewidth]{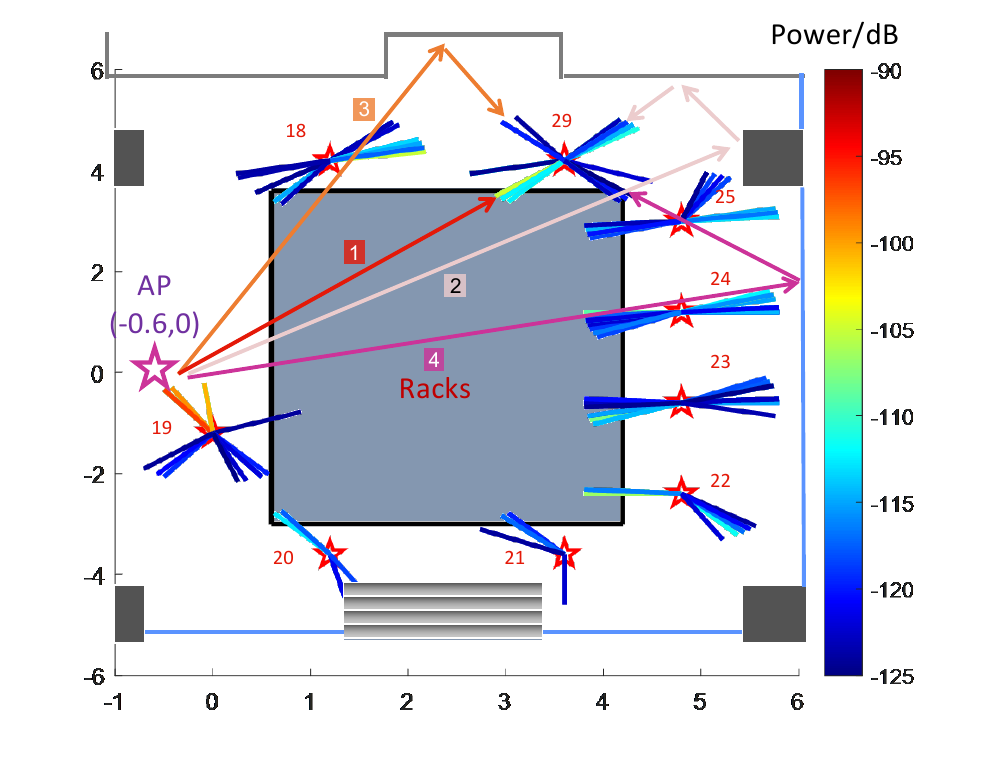}
    }
    \hfill
    \subfloat[]{
        \includegraphics[width=0.83\linewidth]{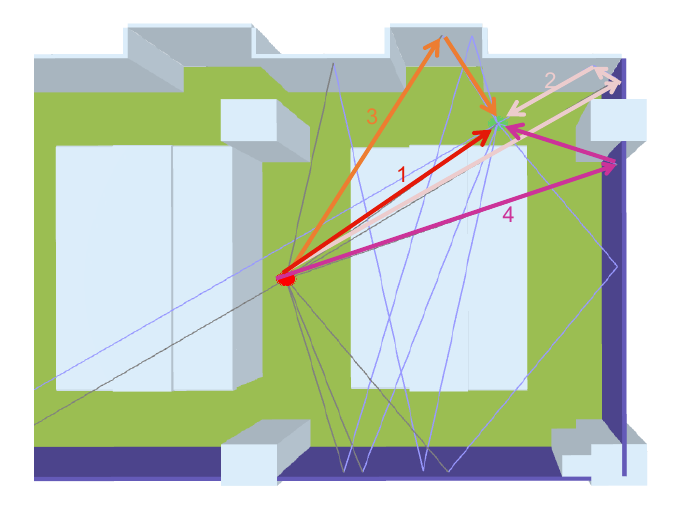}
    }
    \caption{MPC Twins. (a) Physical twin. (b) RT twin.}
    \label{fig:mpcs}
\end{figure}

Fig.~\ref{fig:mpcs} (a) illustrates the spatial distribution of the extracted MPCs projected onto the two-dimensional $x-y$ plane after channel parameter estimation. The figure reveals the spatial characteristics of MPC propagation across $60\times45$ tested Rx locations at the height of $1.7$~m. At most positions, the dominant MPCs originate from the LoS path as well as reflections from the rear-side surfaces. Taking Rx~29 as an example, in addition to the LoS component, several reflected paths are observed, which are primarily attributed to reflections from metallic walls or glass surfaces in the data center environment.
Due to the use of highly directional antennas with narrow beamwidths at the Tx, MPCs arriving from off-boresight directions exhibit relatively low power. As a result, these weak components are suppressed or filtered out during the channel parameter extraction process, leading to a sparse but physically meaningful multipath representation.

Fig.~\ref{fig:mpcs} (b) presents the corresponding coarse RT results. It can be observed that RT is capable of capturing the dominant directions of arrival of MPCs at each Rx location, showing a one-to-one correspondence with the measured MPCs. However, while RT provides a reasonable approximation of the multipath structure at discrete locations, a more refined and spatially continuous representation of the RF field is required for accurate digital twin construction. This motivates the use of the proposed INF-based representation, which is introduced in the following subsection.

\subsection{INF-based AI Twin Construction}

\begin{figure}
    \centering
    \subfloat[]{
        \includegraphics[width=0.86\linewidth]{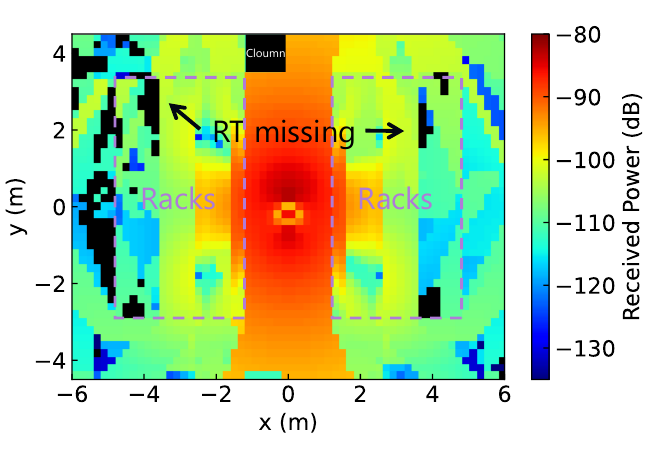}
    }
    \hfill
    \subfloat[]{
        \includegraphics[width=0.86\linewidth]{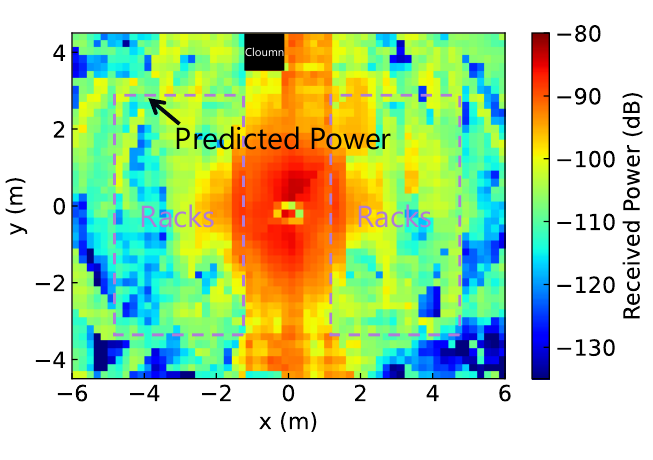}
    }
    \caption{Received power map. (a) RT results. (b) prediction by the AI twin.}
    \label{fig:field}
\end{figure}

Fig.~\ref{fig:field} compares the received power maps predicted by the proposed AI twin and those obtained from the RT twin. As shown in Fig.~\ref{fig:field} (b), the INF-based AI twin produces a spatially continuous received power field, capturing both the dominant LoS region and the complex power variations in NLoS areas. In contrast, the RT result in Fig.~\ref{fig:field} (a) exhibits piecewise-constant patterns and abrupt transitions, which stem the RT twin.
A key observation is that the AI twin demonstrates strong predictive capability in NLoS regions that are not explicitly covered by deterministic ray paths. In these areas, where the RT twin either fails to generate valid propagation paths or significantly underestimates the received power, the INF successfully infers physically consistent signal strength levels by learning from sparse measurement anchors and RT-derived geometric priors. This highlights the advantage of the proposed approach in compensating for missing or weakly modeled multipath contributions in RT, especially in blockage-dominated and reflection-rich environments.
Moreover, the INF representation inherently models the RF propagation as a continuous spatial field rather than a set of discrete evaluation points. This continuity enables smooth power variations across space and avoids artificial discontinuities commonly observed in RT-based maps. Such a continuous RF field is particularly valuable for coverage analysis, interference estimation, and fine-grained digital twin construction, where reliable interpolation between measured or simulated locations is essential.

\begin{figure}[t]
    \centering
    \includegraphics[width=0.88\linewidth]{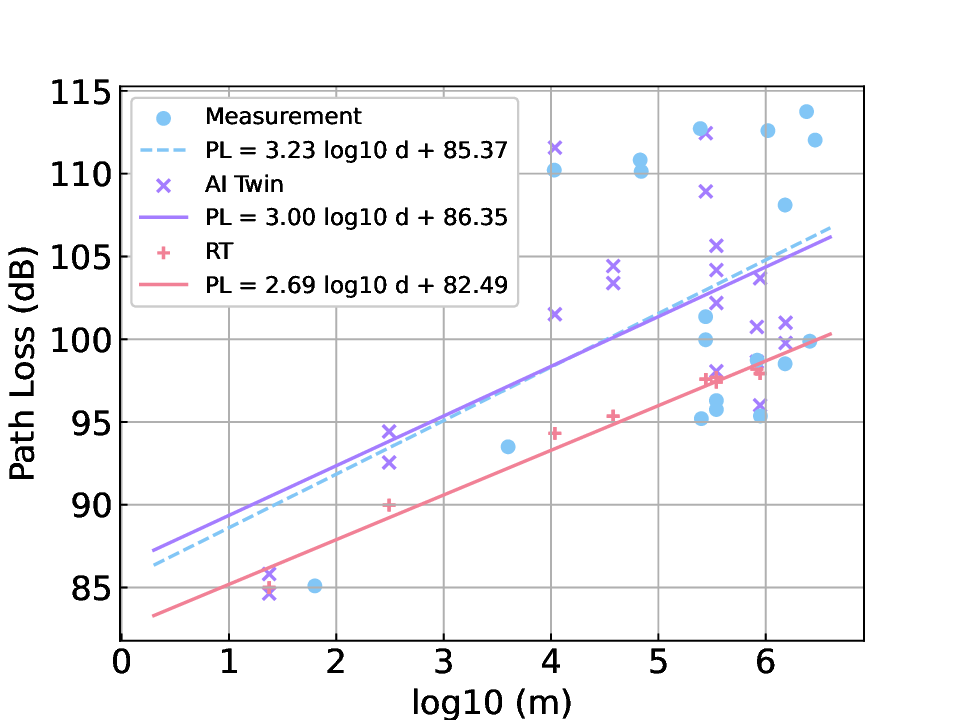}
    \caption{Path loss modeling of measurement, RT twin and AI Twin.}
    \label{fig: plcom}
\end{figure}

As shown in Fig.~\ref{fig: plcom}, the path loss characteristics obtained from the physical twin, RT twin, and the AI Twin are compared as a function of the Tx-Rx separation distance. The physical twin results serve as the ground truth, capturing both LoS and NLoS propagation effects in the data center environment. The RT-based model is able to reproduce the overall distance-dependent trend of path loss, particularly in LoS-dominated regions, but exhibits noticeable deviations in NLoS scenarios due to unmodeled scattering, diffraction, and blockage effects.
By contrast, the proposed AI Twin closely follows the measurement-based path loss across the entire distance range. In particular, in NLoS regions where RT does not work, the AI Twin effectively compensates for RT-missing propagation mechanisms, leading to a more accurate and smoother path loss representation. This demonstrates that, by integrating measurement data with RT-derived structural priors, the AI Twin can provide a reliable and spatially continuous path loss model, which is essential for subsequent coverage and interference analysis.

\subsection{System-level Evaluation Example}
\begin{figure}
    \centering
    \includegraphics[width=0.88\linewidth]{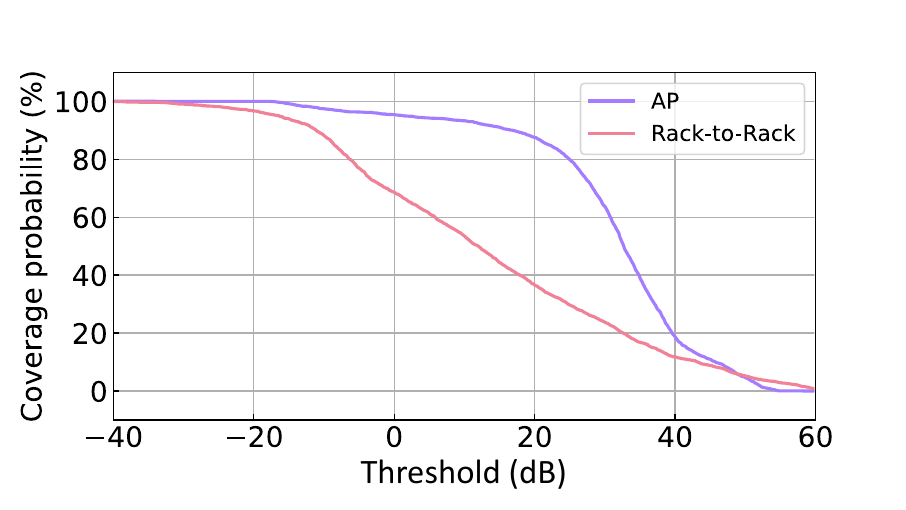}
    \caption{Coverage Probability}
    \label{fig: coverage}
\end{figure}

Fig.~\ref{fig: coverage} illustrates an example of coverage performance at ceiling-mounted AP (Tx~3) locations and at rack-level Tx positions. Taking an SINR threshold of 0 dB as an example, the ceiling-mounted AP is able to provide coverage to more than 95\% of the positions within the data center environment. In contrast, a Txlocated on a single rack covers only approximately 70\% of the area, as inter-rack communication suffers from severe signal degradation caused by dense obstructions and structural blockages in the hall of the data center.
These results indicate that rack-to-rack links are significantly more vulnerable to shadowing and blockage effects than AP-based access links. Through coverage analysis enabled by the digital twin, the impact of deployment height, obstruction density, and network topology on system performance can be quantitatively evaluated. Consequently, the DT provides an effective tool for optimizing AP placement and network architecture, enabling improved coverage, enhanced robustness, and more efficient utilization of wireless resources.

\section{Conclusion}
\label{section: con}
In this paper, a DT framework tailored for THz-WDC is developed. Extensive channel measurements are first conducted using a VNA under both LoS and NLoS scenarios, functioning as the physical twin. 
RT simulations are performed using the Sionna platform and calibrated with measurement data, establishing the RT twin. The RT twin are shown to be largely consistent with the measured multipath characteristics, validating their ability to capture the dominant propagation mechanisms in the data center environment. An RT-conditioned INF–based AI twin is constructed to represent the spatially continuous distributions of received power, angles, and delays. The proposed AI twin further incorporates dedicated calibration for RT-missing NLoS regions, enabling accurate field reconstruction beyond deterministic RT visibility.
Ultimately, leveraging the constructed AI twin, coverage and interference analyses were carried out for both ceiling-mounted AP deployments and rack-level Tx configurations. The resulting coverage maps provide quantitative insights into the impact of deployment strategies and environmental blockage, offering practical guidance for AP placement and network architecture design in THz wireless data centers.

\bibliographystyle{IEEEtran}
\bibliography{DC}

\end{document}